# Neuromuscular and metabolic responses during repeated bouts of loaded downhill walking

Emeric Chalchat[1,2*], Julien Siracusa[1,3], Luis Peñailillo[4], Alexandra Malgoyre[1,3], Cyprien Bourrilhon[1,3], Keyne Charlot[1,3] Vincent Martin[2,5†], and Sebastian Garcia-Vicencio[1,3,6†]

[1] Institut de Recherche Biomédicale des Armées, Bretigny-Sur-Orge, France

[2] Université Clermont Auvergne, AME2P, F-63000, Clermont-Ferrand, France

[3] LBEPS, Univ Evry, IRBA, Université Paris Saclay, 91025 Evry, France

[4] Exercise and Rehabilitation Sciences Institute, School of Physical Therapy, Faculty of Rehabilitation Sciences, Universidad Andres Bello, Santiago 7591538, Chile

[5] Institut Universitaire de France (IUF), Paris, France

[6] Human Motion Analysis, Humanfab, Aix-en-Provence, France

[†]These authors have contributed equally to this work and share last authorship.

***Corresponding author**

Emeric Chalchat, PhD

Institut de Recherche Biomédicale des Armées,

1 place du général Valérie André,

91223 Bretigny-Sur-Orge

France

E-mail: emchalchat@gmail.com

**Short title:** Downhill walking and repeated bout effect

# ABSTRACT

**Introduction:** The aim of this study was to compare *vastus lateralis* (VL) and *rectus femoris* (RF) muscles for their nervous and mechanical adaptations during two bouts of downhill walking (DW) with load carriage performed two weeks apart. Moreover, we investigated cardio-metabolic and perceived exertion responses during both DW bouts.

**Methods:** Seventeen participants performed two 45-min sessions of loaded DW (30% of body mass; slope: -25 %; speed: 4.5 $km \cdot h^{-1}$) separated by two weeks. Rating of perceived exertion (RPE), cost of walking ($C_w$), heart rate (HR), and EMG activity of thigh muscles were assessed during the DW. Muscle shear elastic modulus (µ) of RF and VL were assessed before each exercise bout. Maximal voluntary contraction (MVC) torque was assessed before (PRE), immediately after (POST), 24 and 48h after the two exercise bouts.

**Results:** MVC torque decreased from POST (-23.7±9.2%) to 48h (-19.2±11.9%) after the first exercise (Ex1), whereas it was significantly reduced only at POST (-14.6±11.0%) after the second exercise (Ex2) ($p < 0.001$). RPE (Ex1: 12.3±1.9; Ex2: 10.8±2.0), HR (Ex1: 156±23 bpm; Ex2: 145±25 bpm), $C_w$ (Ex1: 4.5±0.9 $J \cdot m^{-1} \cdot kg^{-1}$; Ex2: 4.1±0.7 $J \cdot m^{-1} \cdot kg^{-1}$) and RF EMG activity (Ex1: 0.071±0.028 mV; Ex2: 0.041±0.014 mV) were significantly decreased during Ex2 compared to Ex1 ($p < 0.01$). RF µ was significantly greater in Ex2 (0.44±0.18) compared to Ex1 (0.56±0.27; $p < 0.001$).

**Conclusion:** The RF muscle displayed specific mechanical and nervous adaptations to repeated DW bouts as compared to VL. Moreover, the muscle adaptations conferred by the first bout of DW could have induced greater exercise efficiency, inducing lesser perceived exertion and cardio-metabolic demand when the same exercise was repeated two weeks later.

**Keywords:** Eccentric exercise; Muscle damage; Fatigue; Muscle stiffness; Elastography; Repeated bout effect

# Introduction

Exercise including unaccustomed eccentric (i.e., lengthening) contractions induces muscle damage that is characterized by a wide range of symptoms such as a prolonged reduction of muscle force and power production (i.e., muscle function) after exercise (1, 2). Moreover, the reduction in muscle function is accompanied by the changes in other indirect markers of exercise-induced muscle damage (EIMD) (e.g., muscle soreness, increases in muscle stiffness and proteins in the blood or reduced range of motion) which can last for several days to weeks after exercise (3). However, when a secondary bout of eccentric exercise is performed, muscle damage is much less, probably due to the neuromuscular (NM) system becomes less susceptible to EIMD (4, 5). This protective effect against muscle damage is widely known as the repeated bout effect (RBE) and has been related to cellular, neural, inflammatory, mechanical adaptations, and extracellular matrix remodeling that might work independently or synergistically to generate muscle protection against subsequent damage (6). Interestingly, Penailillo et al. (7) showed a lower metabolic cost during the second bout of eccentric cycling compared to the first bout performed two weeks before. However, although positive adaptations to repeated eccentric exercise may have a beneficial effect on aerobic performance in healthy and pathological populations (8), the exact mechanisms involved in this adaptation are not fully understood.

Previous studies suggested that neural adaptations may occur in response to a single eccentric exercise (9, 10) or eccentric training (11), and may lead to a reduced energy cost during a repeated bout of eccentric exercise, translating into lower cardio-metabolic responses (7, 9). Specifically, the central nervous system may adjust its recruitment strategies to distribute the mechanical stress over a greater motor unit pool to protect the muscle from damage during a subsequent exercise. Penailillo et al. (9) found lesser EMG activity responses during the second eccentric bout compared to the first performed two weeks before. However, some studies (12-14) failed to find differences in EMG activity between two eccentric exercises performed 1-4 weeks apart. Hyldahl et al. (6) also reported that skeletal muscle extracellular matrix remodeling could increase passive stiffness of the whole muscle the weeks following a bout of eccentric exercise to confer protection against a subsequent damaging exercise bout. As a result, muscles could become more resistant to mechanical strain during the second bout. Consistent with the assumption that the muscle could become more resistant to overstretching, it has been recently shown that mechanical changes (i.e., increase in muscle stiffness) could partly contribute to explain the RBE (i.e., lower fatigue and quicker force-generating capacity recovery) (15). As EIMD could induce an increase in EMG activity during submaximal contractions (16), the protection conferred by mechanical adaptations would translate into a lesser EMG activity increase during a second bout of eccentric exercise at the same workload.

The protection conferred by neural or mechanical adaptations to eccentric exercise may have two complementary effects: it may accelerate the recovery after the damaging exercise and it could modify the NM strategies to limit the development of NM fatigue and muscle damage during the course of the repeated exercise bout. Indeed, it has been shown that NM fatigue was lower immediately after a second bout compared to a first bout (17, 18). Traditionally, the literature on the RBE has essentially focused on the recovery kinetics of symptoms and NM properties. Yet, understanding how the NM system behaves during a repeated bout of eccentric exercise is fundamental for our knowledge and understanding of the RBE phenomenon. Moreover, it has been shown that mechanical and neural adaptations and protection conferred by an eccentric exercise/training vary between individual muscles within a muscle group. Long-term changes in muscle stiffness was different between knee extensor muscles (i.e., *Rectus Femoris* (RF) vs. *Vastus Lateralis* (VL)) after a single eccentric exercise (15). It is possible that RF and VL are heterogeneously stretched during the DW exercise. Thus, as RF $\mu$ was more affected than VL $\mu$ after a DW, it could be that the mechanical stress induced by the walking position during loaded DW placed greater strain on the RF than the VL, leading to different adaptations (6). Indeed, the trunk position and the load distribution during loaded

DW favors hip extension, which induces a relatively greater lengthening of RF than VL due to their biomechanical properties (i.e., mono- and bi-articular muscles). Consequently, RF may exercise at a longer muscle length and could be more susceptible to muscle damage and adaptations. Moreover, changes in maximal voluntary EMG activity were different between plantar flexors muscles (i.e., *Soleus* vs. *Gastrocnemius Medialis*) after a 4-week eccentric training (11). It is therefore suggested that changes in neuromechanical coupling during the course of two eccentric exercises performed several weeks apart may differ between muscles within a muscle group. However, it is not known whether the mechanical adaptations to eccentric exercise/training could differently affect muscle activation, an indicator of NM efficiency and fatigue, during an ecological exercise task (e.g., locomotion). Therefore, the aim of this study was to investigate how VL and RF activity adapt during two downhill walking (DW) with load carriage performed two weeks apart, a model known to induce mechanical adaptations on RF but not on VL (15). We hypothesized that neural and mechanical adaptations conferred by the repeated bout effect would be greater in RF than VL due to exercise model specificity. Moreover, we hypothesized a lesser cardio-metabolic demand and perceived exertion responses during the second exercise bout compared to the first.

# Materials and methods

## Participants

Seventeen young males (age: 31.6 ± 5.7 years, body mass: 77.8 ± 11.1 kg, height: 180 ± 6 cm, body mass index: 24.0 ± 2.4 $kg \cdot m^{-2}$, and fat mass: 15.6 ± 5.3 %) were fully informed of the experimental procedures, gave their written informed consent and agreed to participate in this study. The sample size was calculated on the basis of an α level of 0.05 and a power (1 - α) of 0.8, based on the difference in EMG activity between initial and secondary bouts of eccentric exercise reported previously (9). Participants performed regular physical activity (between 3 and 8 $h \cdot w^{-1}$), had no recent history of muscular, joint, or bone disorders, and did not take any medication or supplementation that could affect NM responses. None of them experienced strenuous mountain trekking and/or DW and none had been involved in a resistance-training program in the past 6 months. Participants were instructed not to perform unaccustomed activities and any interventions that may interfere with recovery such as massage, icing, nutritional supplementation, and anti-inflammatory drugs during the experimental period. Each participant performed a familiarization session, consisting of a complete medical examination, including the collection of anthropometric data and complete familiarization with the experimental measurements. No DW was performed during the familiarization session to avoid any adaptation related to the repeated bout effect. The study was conducted in accordance with the Declaration of Helsinki (19) and was approved by the Regional Ethics Committee (CPP Ile-de-France 8, France, registration number: 2019-A01210-57, SMILE).

## Study design

A schematic representation of the protocol is shown in Figure 1. Participants performed two 45-min DW bouts on a treadmill (pulsar 3p 4.0, HP cosmos, Warwickshire, UK) separated by two weeks at a gradient of -25 % while carrying a load equivalent to 30 % of the body mass (carrying: a weighted vest: 10 % of the body mass; a backpack: 20 % of their body mass). Each participant performed DW bouts at the same time of day (in the morning) to control for within-participant diurnal variation. Before DW exercises, participants performed a 5-min warm-up consisting in walking while carrying loads with progressive increases of velocity (from 2.5 to 4.5 $km \cdot h^{-1}$) and gradient (from 0 to -25 %). The exercise started when the treadmill gradient was set to -25 % and velocity to 4.5 $km \cdot h^{-1}$ and the participants were instructed to walk at their own preferred stride length and frequency. Five participants were not able to finish the 45-min walking bout and performed the first exercise bout to exhaustion (30.2 ± 8.6 min). These participants performed the same exercise duration during the second exercise.

Surface EMG amplitude of VL, RF and *biceps femoris* (BF), spatio-temporal gait parameters (number of steps, and stance, and swing duration), metabolic variables including heart rate (HR), oxygen consumption ($VO_2$), and rating of perceived exertion (RPE) were obtained during walking. Maximal voluntary contraction (MVC) torque and quadriceps muscle soreness were assessed before (PRE), immediately after (POST), and 24 and 48 h after the two DW bouts, as indirect markers of muscle damage. Resting shear elastic modulus (μ) of the VL and the RF were assessed by shear wave elastography (SWE) before each exercise (i.e., 14 days apart), as markers of muscle stiffness adaptations.

## Measurements

### Before and after exercise

#### *Maximal voluntary contraction torque*

MVC torque was measured using an isokinetic dynamometer (Cybex Norm, Lumex, Ronkonkoma, NY, United States). Participants were comfortably positioned on an adjustable chair with the hip joint flexed at 70° (0° = neutral supine position). The axis of rotation of the dynamometer was aligned with the femoral condyles of the femur and the lever arm was attached 1-2 cm above the malleolus with a Velcro strap. All measurements were taken from the participant's right leg (knee angle = 90°; 0° = knee fully extended). Torque data were corrected for gravity, digitized and exported at a rate of 2 kHz to an external analog-to-digital converter (Powerlab 16/35; ADInstruments, New South Wales, Australia) driven by the Labchart pro 8.1 software (ADInstruments, New South Wales, Australia). During each 5-s MVC, the participants were instructed to grip the lateral handles of the seat in order to stabilize the pelvis and were strongly encouraged by the investigator to push as fast and hard as possible. Two MVC were performed with a 2-min rest period between each contraction. MVC torque was determined as the peak torque reached during the maximal effort and the higher value was used for further analyses.

### *Muscle soreness*

The magnitude of muscle soreness of the whole quadriceps was assessed using a visual analog scale, consisting of a 100-mm line representing "no pain" at one end (0 mm), and "very, very painful" at the other (100 mm), while performing a single squat over a 90°of knee range of motion.

An algometer (NOD, OT Bioelettronica, Turino, Italy) with a rigid plastic rod covered by a rubber surface with an area of 1.0 $cm^2$ and a scale ranging from 0 to 500 kPa was used to measure muscle pain pressure threshold (PPT). First, the procedures were explained clearly to the subject. The participants were evaluated in a sitting position with the knee placed at 90° of flexion (0° = knee fully extended), in a relaxed state. The compression pressure was applied on the mid portion of the VL and RF muscles. The pressure was applied vertically, and it increased at a constant rate of 50 kPa per second. Participants were asked to say "stop" when feeling muscle pain or discomfort. Two repetitive measurements were performed for each muscle at an interval of 60 s.

### *Resting shear elastic modulus*

An ultrafast ultrasound scanner (Aixplorer version 12.2; Supersonic Imagine, Aix-en-Provence, France) coupled with a linear transducer array (SuperLinear 15–4; Supersonic Imagine, Aix-en-Provence, France) was used in both SWE (musculoskeletal preset, penetration, no persistence) and research modes, as previously proposed (20). The B-mode ultrasound was first set to determine the optimal transducer location and maximize the alignment between the transducer and the direction of the RF and VL muscle fascicles. Transducer alignment was considered correct when muscle fascicles and aponeurosis could be delineated across the image without interruption. The transducer was then fixed at 50 % of the total muscle length respectively using a dynamic probe fixation device (with 360° adjustments, USONO, Eindhoven, Netherlands) placed over the skin, which was coated with a water-soluble transmission gel (Aquasonic, Parker laboratory, Fairfield, NJ, United States) to ensure acoustic coupling. This fixation allows to avoid any movement which could affect the position and the orientation of the probe, and avoid excessive pressure applied to the muscle. The position of the probe was marked on the skin using a permanent marker to ensure the same positioning along the experiment.

A fixed-size rectangular region of interest (ROI), i.e., the region in which shear-wave propagation was analyzed within the muscle, was placed in the middle of the B-mode image below the superficial aponeurosis within the VL and RF muscle belly. The position of the ROI was carefully located on the B-mode image during the familiarization session to be sure to keep the same ROI at each measurement point (before each exercise). Five-second SWE sequences (frame rate: 1–2 Hz) were performed at rest for VL and RF muscles with the knee joint positioned at 90° and 120° (0° = knee fully extended).

A two-dimensional real-time shear wave velocity (in $m \cdot s^{-1}$) color map with a spatial resolution of 1 × 1 mm was obtained, as previously described (15, 21, 22). The shear wave velocity raw data was then transferred to a workstation, converted to shear elastic modulus (μ) and analyzed using a MATLAB script developed in our laboratory (MathWorks, Natick, MA, United States). The μ was obtained as follows:

$$\mu = \rho \cdot Vs^2$$

where ρ is the muscle density (1,000 $kg \cdot m^{-3}$) and Vs is the shear wave velocity (in $m \cdot s^{-1}$). The μ values were averaged over the ROI (a ~36 × 114 matrix) without the empty (0 $m \cdot s^{-1}$) and saturated (16.32 $m \cdot s^{-1}$) values, and the average of 5-8 consecutive reconstructed images from raw data (available for Aixplorer version 12.2 with research pack) was used for subsequent analyses. As previously described by Lacourpaille et al. (23), the slope of the relationship between the change in μ and the knee joint angle of 90° and 120° at each time point (before each exercise) was calculated to compare VL and RF μ, regardless of their relative length.

**During exercises**

***Spatio-temporal parameters of DW***

Participants were equipped with four tri-axial inertial sensors with EMG all-in-one system (Trigno Avanti System, Delsys, Boston, Massachusetts, USA) to assess the kinematic and synchronize it to EMG data from RF, VL and BF muscles. Sensors were attached with tape and placed above and posterior to the right and left ankles (over the *soleus* muscle) and on the proximal third of the leg over the *tibialis anterior* muscle. The sampling frequency was set to 143 Hz and 3D signals were synchronized and continuously recorded over the entire DW exercise using an external analog-to-digital converter (Powerlab 16/35; ADInstruments, New South Wales, Australia) driven by the LabChart pro 8.1 software (ADInstruments, New South Wales, Australia). Signals were then filtered using a second-order zero-lag bidirectional Butterworth filter (cutoff of 10 Hz) (24). Primary kinematic parameters, including 3D angular velocity and 3D accelerations, were used to label automatically each gait phase from foot switches (foot contact and foot off). The gait event was defined as 0 to 100 % from the first left heel contact to the next one. The heel contact was identified as the negative peaks of medial-lateral angular velocity obtained from the gyroscope occurring in heel contact before the instant of the minimum peak of anterior-posterior acceleration. Then, four phases during stance were defined: loading response, midstance, terminal stance, and pre-swing. Since foot switches provide no information during the swing, the swing phase was defined as the total duration of the swing, i.e., not defined into any subphases. To confirm the first heel contact, a video camera synchronized to signals (video capture module) driven by the LabChart Pro 8.1 software (ADInstruments, New South Wales, Australia) was used. Stance, and swing duration were expressed as a mean for each exercise. A semi-automated MATLAB2019a script developed in our laboratory (MathWorks, Natick, MA, United States) was used for the entire analysis. The number of steps and the mean of the entire exercise for stance and swing duration were assessed to test if the walking kinematics were different between the first (Ex1) and the second (Ex2) exercise.

***EMG activity***

The EMG signals of the VL, RF, and BF muscles were recorded during DW exercises using active double-differential electrodes with a wireless system (Trigno Avanti System, Delsys, Boston, Massachusetts, USA). The recording electrodes were taped lengthwise to the skin, as recommended by SENIAM (25). The skin was prepared by shaving, gently abrading with sandpaper, and cleaning with alcohol. The position of the EMG electrodes was determined according to SENIAM recommendations (25) and was marked on the skin using a permanent marker to ensure the same positioning along the experiment. The sampling frequency was set at 2000 Hz and the EMG data was passed through a fourth-order zero-lag Butterworth filter using 20-450 Hz digital band-pass filter in order to eliminate noise. EMG data was then smoothed using a root mean square algorithm with a 10-ms window to produce a linear envelope from which the absolute average peak amplitude (mV) of the stance (~0-35 % of the step duration: loading response) and swing (~60-100 % of the step duration) phases was calculated (Figure 2).

***Cost of walking and heart rate***

Breath-by-breath $VO_2$ and $VCO_2$ were measured using a metabolic cart (Quark CPET; Cosmed, Rome, Italy). HR was monitored by a three-channel electrocardiogram consisting in a chest belt (Custo belt, Custo med GmbH, Ottobreunn, Germany) connected to a transmitter (Custo guard, Custo med GmbH, Ottobreunn, Germany). $VO_2$, $VCO_2$ and HR were continuously recorded during each 45-min DW bouts and before warm-up (i.e., at rest in the standing position while carrying loads for 3 min).

Energy expenditure (EE) was calculated from $VO_2$ and $VCO_2$ (in $L \cdot min^{-1}$) using the Jeukendrup and Wallis equation (26):

$$EE = (0.575 \cdot VO_2) + (4.435 \cdot VCO_2)$$

EE (in kcal·min$^{-1}$) was then converted in J·min$^{-1}$. The cost of walking ($C_w$ in J·m$^{-1}$·kg$^{-1}$) representing the energy required to walk 1 m normalized to kg of body mass was then calculated using this equation (27):

$$C_w = \frac{EE_{DW} - EE_{rest}}{75 \cdot BM}$$

where $EE_{DW}$ and $EE_{rest}$ are the EE in J·min$^{-1}$ during the DW and the 3-min rest period, respectively, 75 the distance (in m) walked in 1 min, and BM the body mass (in kg).

### *Rating of perceived exertion*

RPE measurements were taken using Borg's 6–20 scale at 1, 5, 15, 25, 35, 40, and 45 min of exercise.

### *RPE, $C_w$, HR and EMG activity synchronization*

A mean of $C_w$, HR, and RF and VL EMG activity were calculated for every 5 % of the total DW duration. RPE data were linearly interpolated every 5 % of the total DW duration. RF and VL EMG activity were analyzed during the stance phase (i.e., ~0-35 % of the step duration; Figure 2) because these muscles are the most active in response to body weight support during this phase (28). RF and VL EMG activity were also analyzed during the swing phase (i.e., ~35-100 % of the step duration) to compare result with those of Penailillo et al. (9) observed during the recovery phase of eccentric cycling. BF EMG activity was analyzed during ~60-100% of the step duration phase (i.e., second part of the swing phase) because this muscle is the most active during this phase. BF EMG activity was also analyzed during the stance phase (i.e., ~0-35 % of the step duration) to evaluate any difference in co-activation.

## Statistical analysis

The data were screened for normality of the distribution and homogeneity of variances using the Shapiro–Wilk normality and Levene´s tests, respectively. MVC torque, RPE, $C_w$,

HR, EMG activity data were analyzed by two-way ANOVA (time × exercise) with repeated measures for paired values. When an interaction between time and exercise was found by the two-way ANOVA, a Bonferroni post-hoc test was applied to test for differences between the means. For muscle soreness, because of the absence of a normal distribution according to the Shapiro-Wilk test, we performed a non-parametric Friedman analysis of variance to assess the main effect of time on muscle soreness. Dunn's multiple comparisons were used to identify which time points were significantly different to baseline for these variables. Statistical differences between matched time points (i.e., Ex1 vs. Ex2) were assessed using Mann-Whitney U test. Each P value was adjusted to account for multiple comparisons. Between-exercises (i.e., Ex 1 vs. Ex 2) differences of RF and VL μ, the number of steps, and mean values of the entire exercise for stance and swing duration were analyzed using a paired-sample t-test. Statistical tests were performed using the Statistica 8.0 sofware (StatSoft, Inc, USA). The results are presented as the means ± SD. Significance was defined as $P < 0.05$.

# Results

## Before and after exercise

### *MVC torque*

Absolute PRE values of MVC torque were not different between Ex1 and Ex2 (315 ± 50 Nm vs. 311 ± 70 Nm, $p = 0.81$). A significant interaction (time × exercise; $p < 0.001$) was found for MVC torque (Figure 3). MVC torque decreased until 48 h after Ex1, whereas it was significantly lesser only at POST after the Ex2 (Figure 3).

### *Muscle soreness*

A significant effect of time was found for muscle soreness ($p < 0.001$). Muscle soreness increased until 48 h after both Ex1 and Ex2 (Figure 4A). However, muscle soreness was significantly lesser after Ex2 than after Ex1 (Figure 4A).

A significant effect of time was found for RF and VL PPT ($p < 0.001$), but no significant difference was found between RF and VL PPT ($p = 0.09$). RF PPT decreased only 48 h after Ex1 (Figure 4B), whereas VL PPT decreased from 24 h to 48 h after Ex1 (Figure 4C). Moreover, PPT was significantly higher after Ex2 than Ex1 at 48 h for RF (Figure 4B) and from 24 h to 48 h for VL (Figure 4C).

***Resting shear elastic modulus***

RF µ PRE value was significantly greater before Ex2 compared to Ex1 (Figure 5). Such a difference was not observed for VL µ PRE values between Ex1 and Ex2 (Figure 5)

**During exercises**

***Spatio-temporal parameters of DW***

The number of steps (Ex1: 2238 ± 668 steps; Ex2: 2147 ± 698 steps; $p = 0.12$) and mean of the entire exercise for the stance (Ex1: 0.66 ± 0.05 s; Ex2: 0.66 ± 0.05 s; $p = 0.70$) and swing (Ex1: 0.43 ± 0.03 s; Ex2: 0.44 ± 0.03 s; $p = 0.24$) durations were not different between Ex1 and Ex2.

***EMG activity***

A significant interaction effect (time × exercise; $p < 0.001$) was found for RF EMG activity during the stance phase (Figure 6A). RF EMG activity increased later and started to plateau earlier in the Ex2 compared the Ex1 (Figure 6A). RF EMG activity was significantly lower during the Ex2 compared to Ex1 during the entire exercise (Figure 6A). During the swing phase, no significant interaction (time × exercise; $p = 0.99$) or time effect ($p = 0.90$) was found for RF EMG activity. However, a significant effect of exercise was found: RF EMG activity was, on average, lower during Ex2 than Ex 1 (Ex1: 0.023 ± 0.007 mV; Ex2: 0.016 ± 0.004 mV; $p < 0.001$).

No significant interaction (time × exercise; $p = 0.54$) or exercise effect ($p = 0.35$) was found for VL EMG activity during the stance phase (Figure 6B). However, a significant effect of time ($p < 0.001$) was found: VL EMG activity increased and started to plateau at 35-40 % of the total exercise duration (Figure 6B). During the swing phase, no significant interaction (time × exercise; $p = 0.99$) or exercise ($p = 0.24$) and time ($p = 0.07$) effect was found for VL EMG activity.

No significant interaction (time × exercise; $p = 0.20$), exercise ($p = 0.91$) or time ($p = 0.69$) effect was found for BF EMG activity during the swing phase (Figure 6C). Similarly, no significant interaction (time × exercise; $p = 0.96$), exercise ($p = 0.10$) or time ($p = 0.90$) effect was found for BF EMG activity during the stance phase.

***Cost of walking***

A significant interaction (time × exercise; $p = 0.002$) was found for $C_w$ (Figure 7A). $C_w$ increased throughout the exercise, except between 5-10 % of total exercise duration, for both exercises (Figure 7A). $C_w$ was significantly lower during Ex2 compared to the Ex1, except at the beginning of the exercise (5-10 % and 15-30 % of total exercise duration; Figure 7A). $C_w$ stabilized later during Ex1 than Ex 2 (80-85 % vs. 60-65 % of total exercise duration for Ex1 and Ex2 respectively (Figure 7A)).

***Heart rate***

A significant interaction (time × exercise; $p = 0.009$) was found for HR (Figure 7B). HR increased throughout the exercise for both exercises, except between 5-10 % of total exercise duration, for the Ex1 (Figure 7B). HR was significantly lower during Ex2 compared to Ex1 (Figure 7B).

***Rating of perceived exertion***

A significant interaction (time × exercise; $p < 0.001$) was found for RPE (Figure 7C). RPE increased throughout the exercise for both exercises (Figure 7C). RPE was significantly lower during Ex2 compared to Ex1 (Figure 7C).

# Discussion

The present study compared perceived exertion, cardio-metabolic (i.e., Cw and HR) and neuromechanical coupling (i.e., RF and VL EMG activity and mechanical properties) changes during two loaded DW bouts (Ex1 and Ex2) performed two weeks apart. The results support the hypothesis that neural and mechanical adaptations conferred by the repeated bout effect would be greater in RF than VL due to exercise model specificity. Moreover, the results support the hypothesis that cardio-metabolic demand and perceived exertion responses were lower throughout the second exercise bout compared to the first.

The two DW bouts were performed with the same kinematics, as evidenced by the similarity of the number of steps, and mean stance and swing duration. However, the RPE, HR, $C_W$, and RF EMG activity were significantly lower during Ex1 and Ex2. The results showed that RF EMG activity was lower and increased less during Ex2 compared to Ex1, whereas no significant difference in EMG activity was found for VL. These changes in RF EMG activity between Ex2 and Ex1 were observed with a concomitant increase in RF stiffness (i.e., μ) observed before Ex2 (i.e. 14 days after Ex1). Taken together, these results support the hypothesis that Ex1 conferred adaptations protecting specifically the RF muscle (i.e., increase in muscle stiffness and lower EMG activity in Ex2). These adaptations could be involved in the better walking performance (i.e., better walking efficiency and lower HR) and the lower perceived exertion observed during Ex2 than Ex1.

In line with previous studies (29, 30) that compared two loaded walking bouts (slope: −28%; velocity: 5 km·h$^{-1}$; load: 10 % of body mass; 1 week apart), the results of the present study showed a protective effect of Ex1 on Ex2 (i.e., RBE). Similar to these previous studies, we found a prolonged MVC torque loss (i.e., > 24 h after the exercise) of the knee extensors after Ex1 (~20-25 %; moderate muscle damage (2, 3)) whereas MVC torque had already recovered 24 h after Ex2. Moreover, muscle soreness, another marker of EIMD (1, 3), was less affected after Ex2 compared to Ex1. Hyldahl et al. (6) explained this phenomenon as a protective effect against muscle damage related to cellular, neural, inflammatory and mechanical adaptations.

In the present study, the protective effect was also illustrated by a lower perceived exertion and a better walking performance (i.e., lower RPE and $C_w$) during Ex2 compared to Ex1. These results are partially in line with Penailillo et al. (7) who found a lower HR after a second eccentric cycling bout compared to the same exercise performed two weeks before, but no significant differences in RPE and $VO_2$ between these two exercises. Penailillo et al. (7) also found lower blood lactate levels during the second bout compared to the first bout, suggesting lower metabolic stress during the repeat eccentric exercise bout. These are interesting results demonstrating that positive adaptations associated to eccentrically-biased exercises also induce changes in the cardio-metabolic responses, which may determine aerobic performance in healthy and pathological populations. However, the underlying mechanisms inducing these adaptations were unknown until now.

The results of the present study revealed that RF EMG activity was lower during Ex2 compared to Ex1 during both the stance (i.e., eccentric phase) and swing phases, whereas no differences in VL EMG activity was found between Ex1 and Ex2 regardless of the phase (i.e., stance or swing) investigated. This is not consistent with Penailillo et al. (9) who found a more efficient motor pattern (i.e., lower VL EMG activity) of the recovery phase of a cycle revolution during a secondary eccentric cycling bout compared to the initial bout. These inconsistencies could be due to the loaded DW, inducing specific adaptations on the RF (15) and the fact that the type of eccentric exercise could differently affect individual muscles within a muscle group (31). This is specifically relevant considering that loaded DW favors hip extension due to the trunk position and the load distribution, and that RF is a bi-articular muscle in comparison to VL. RF would exercise at a longer muscle length and would be more susceptible to muscle damage and adaptations. Indeed, it has been shown that exercising at a greater muscle length induces more muscle damage and greater protection against EIMD (i.e., greater adaptations) (32). The results of the present study are not aligned with some studies that found no differences in EMG activity between two eccentric exercises (knee flexors, elbow flexors, plantar flexors) performed 1-4 weeks apart (12-14). As these authors found no mechanical changes (12) or no evidence for mechanical changes (13, 14) between the first and the second eccentric exercise, it could be that mechanical and neural adaptation are related. For instance, Penailillo et al. (9) found both mechanical (i.e., lesser fascicle elongation) and neural (i.e., lesser EMG activity) adaptations during the second eccentric bout compared to the first performed two weeks before. In the present study, it should be noted that EMG activity was lower during Ex2 compared to Ex1 only for RF, which is the only muscle mechanically adapted (i.e., long-term increase in μ) to Ex1. Indeed, VL μ was not different between Ex1 and Ex2, whereas BF was not eccentrically activated during the DW. However, it is difficult to know if the lower RF EMG activity observed at the beginning of Ex2 compared to Ex1 is due to better neural strategy (10, 11) or due to better mechanical efficiency. Lemire et al. (33) showed that lower limb musculotendinous stiffness was a significant predictor of downhill running performance, whereas Saunders et al. (34) suggested that an increase in muscle stiffness could positively impact functional submaximal tasks that involve stretch-shortening cycle such as walking or running. The results of the current study seem to support this hypothesis.

RF EMG activity increased to a smaller extent during Ex2 compared to Ex1, as we found that RF EMG activity started to plateau earlier in Ex2 (15-20 % of total exercise duration) compared to Ex1 (45-50 % of total exercise duration). The increase in RF EMG activity throughout Ex2 compared to Ex1 was likely due to lower NM fatigue during Ex2 (35). This lower increase in RF EMG activity (i.e., lower NM fatigue) could be explained by a more efficient motor recruitment pattern and/or by a more efficient force transmission due to mechanical adaptation (i.e. stiffer muscle) (34). This greater efficiency could translate into a lesser cardio-metabolic demand and perceived exertion during Ex2. Changes in motor recruitment pattern could be explained by different mechanisms such as (I) a shift in motor unit recruitment toward low-threshold motor units during the second bout (6), (II) an enhanced synchronization of motor units (6), (III) a lower antagonist muscle activation (36) although no significant difference in BF EMG activity during the stance phase in the present study, and (IV) an attenuation of supra-spinal fatigue (18). Moreover, Hyldahl et al. (6) reported that muscle stiffness might increase by skeletal muscle extracellular matrix remodeling following eccentric exercise, which may protect muscle from damage during a subsequent eccentric exercise. It could be that the greater muscle stiffness of RF before Ex2 (i.e., 14 days after Ex1) limited the sarcolemma disruption during Ex2 compared to Ex1 due to more robust muscle and/or connective tissue, limiting NM fatigue. The lower NM fatigue (i.e. lower increase in RF EMG activity) observed in Ex2 of the present study seems to translate into lower deterioration of walking performance, as $C_w$ started to plateau earlier in Ex2 (60-65 % of the entire exercise duration) than in Ex1 (80-85 % of the entire exercise duration).

These findings indicate that a single bout of DW may be a good strategy to improve the walking efficiency and limit the deterioration of walking performance during subsequent DW bouts. This modality of eccentric exercise could be implemented in clinical setups to induce neuromuscular adaptations. Specifically, adjusting load carriage would be a valuable alternative to walking speed modulation or classical high-intensity resistance exercise. However, the value of this application still needs to be evidenced. Furthermore, it is not known if the adaptations to a single bout of DW could be transferred to other activities.

# Conclusion

Protective adaptations conferred by the first bout of DW induced a lower perceived exertion and cardio-metabolic responses when the same exercise was repeated two weeks later (i.e., better walking efficiency). Neural and mechanical adaptations could explain these findings, especially from RF muscle adaptations (i.e., decreased EMG activity and increased muscle stiffness). Indeed, no adaptation was found on VL after a single DW bout, likely due to the specificity of DW, probably due to the walking position in relationship to the treadmill. Further research is needed to investigate factors explaining the relationship between the type of eccentric exercise and the muscle-specific adaptations.

# Acknowledgments

We would like to thank Benoît Lepetit, Stéphane Baugé, Phillippe Colin, Caroline Dussault, Stéphanie Bourdon, Pierre-Emmanuel Tardo-Dino, Walid Bouaziz and Olivier Nespoulous for their technical and/or medical support.

No support was provided for this study by any manufacturer of the instruments used. Additionally, the results of the present study do not represent an endorsement of any product by the authors or the ACSM. No conflicts of interest, financial or otherwise, are declared by the authors. Further, the authors declare that the results of the study are presented clearly, honestly, and without fabrication, falsification, or inappropriate data manipulation.

# Figure legends

**Figure 1. Design of the study.** Participants performed 2 bouts of downhill walking on a treadmill for 45 min while carrying a load equivalent to 30% of their body mass (treadmill gradient: -25 %, speed: 4.5 km·h-1) separated by 14 days (top). During exercises, rating of perceived exertion (red), gas exchanges (green), heart rate (blue), muscle activity (orange) of the *Rectus femoris* (1), *Vastus lateralis* (2) and *Biceps femoris* (3), and the spatio-temporal parameters (red) were obtained during walking.

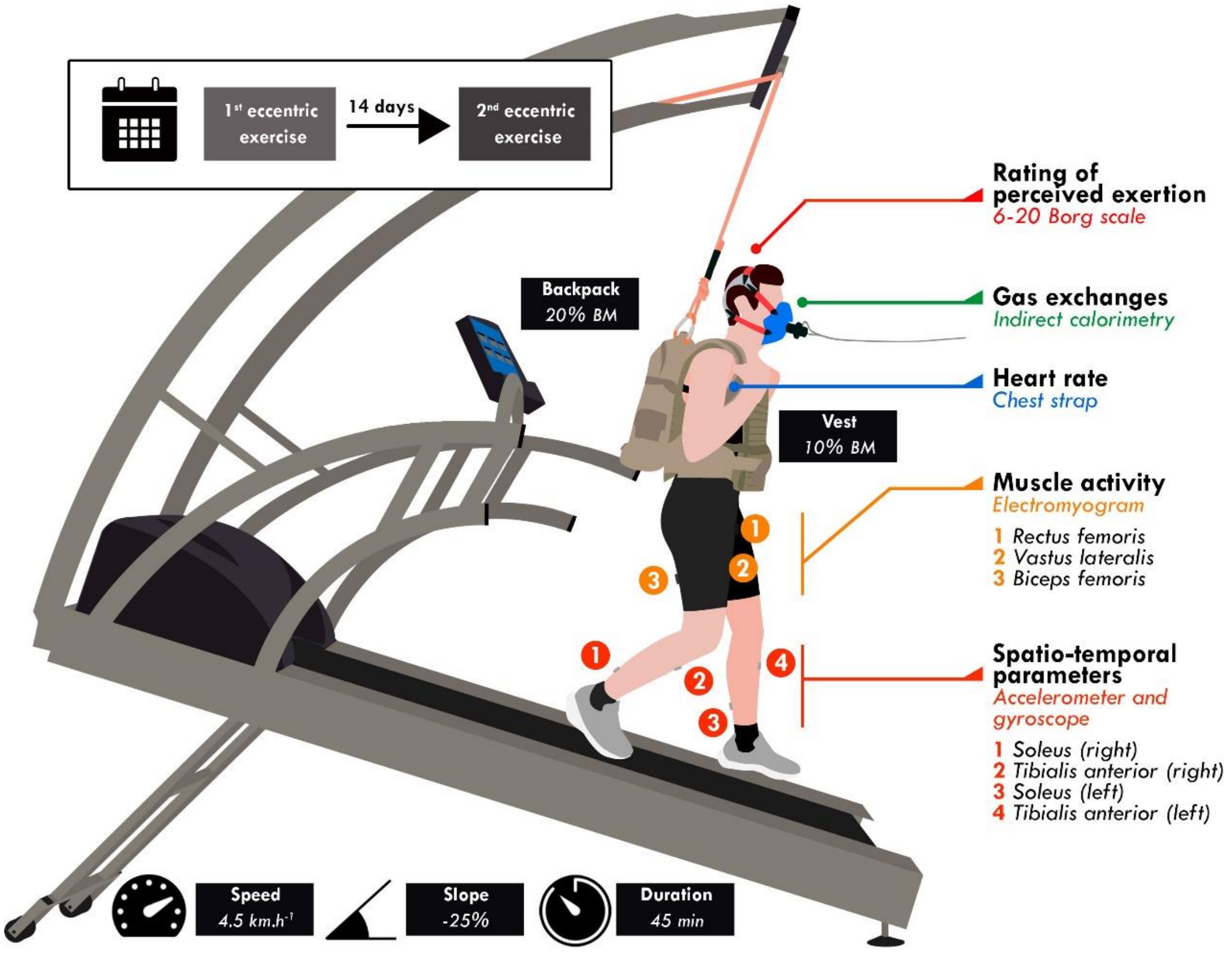

**Figure 2. Mean ± SD of (A) *Rectus Femoris* (RF), (B*) Vastus Lateralis,* and (C) *Biceps Femoris* EMG activity during a step.**

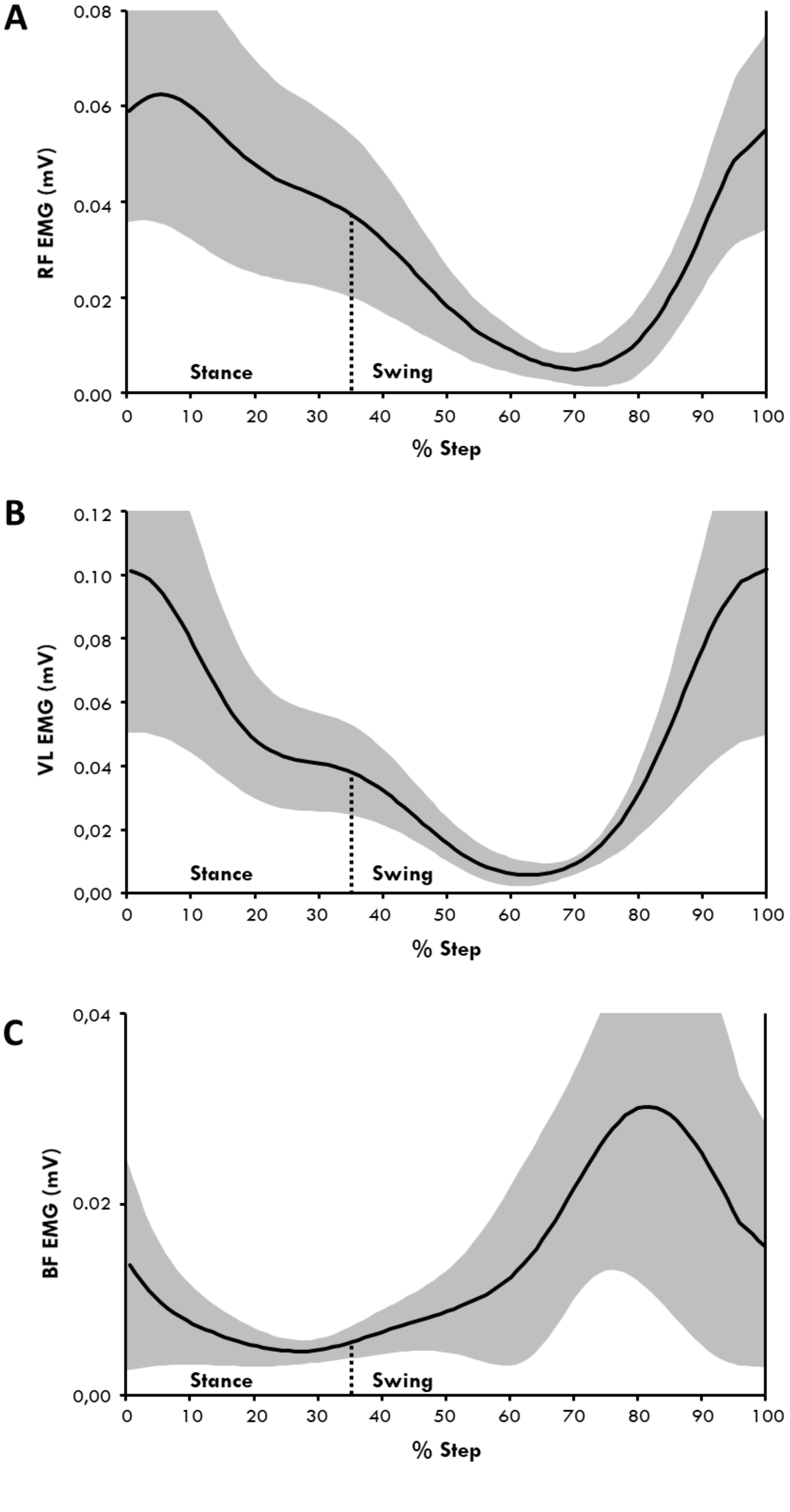

**Figure 3. Time course of changes in maximal voluntary contraction (MVC) torque from PRE to 24 h and 48 h for both exercise bout 1 and exercise bout 2.** Significantly different from the PRE value: *** ($p < 0.001$); significant difference between exercise bout 1 and exercise bout 2: $$$ ($p < 0.01$).

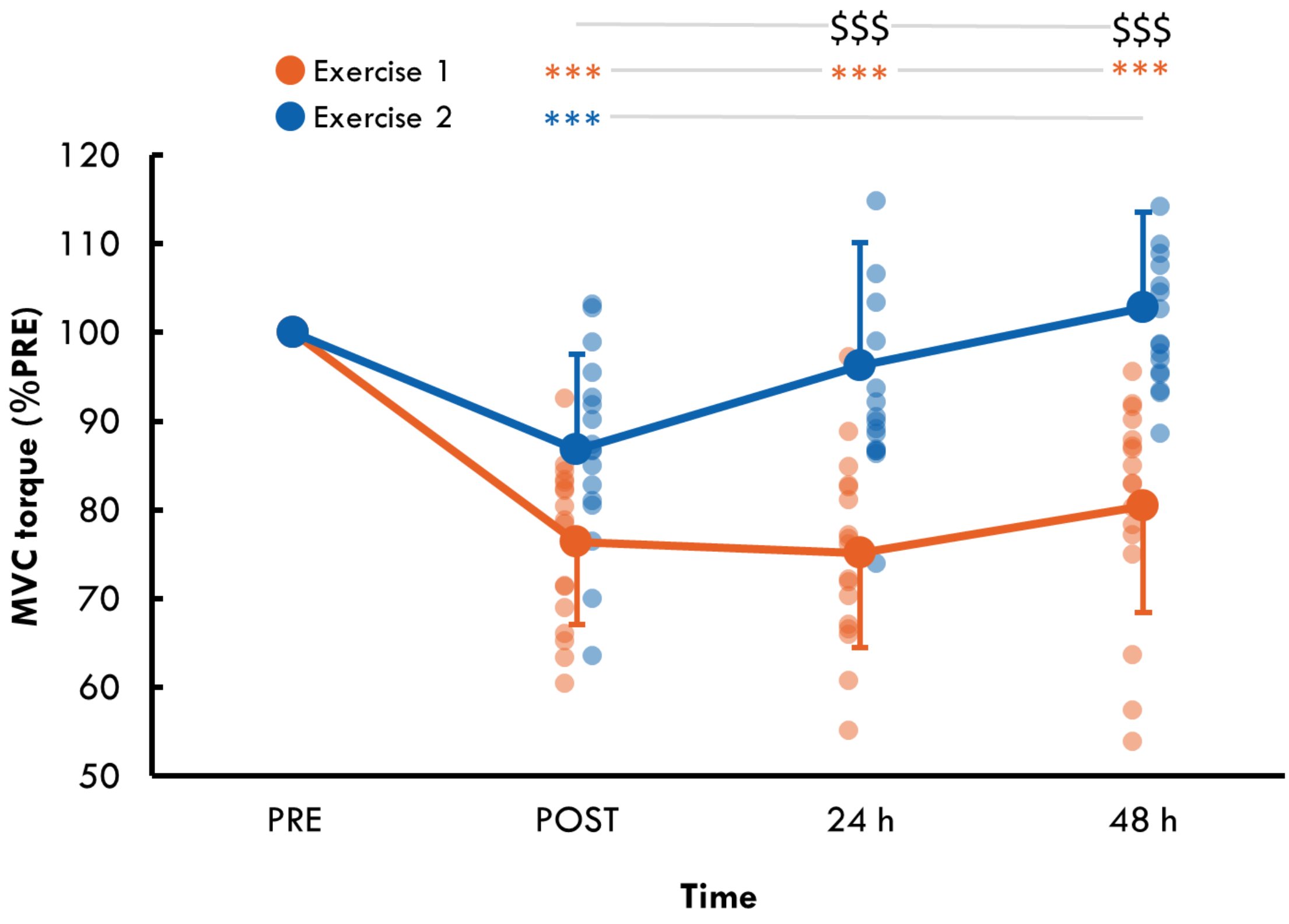

**Figure 4. Time course of (A) muscle soreness, and pain pressure threshold (PPT) for (B) *rectus femoris* (RF) and (C) *vastus lateralis* (VL) for exercise 1 and exercise 2.** Significantly different from PRE value: *: $p < 0.05$, **: $p < 0.01$, ***: $p < 0.001$. Significant difference between Ex1 and Ex2: $: $p < 0.05$, $$: $p < 0.01$.

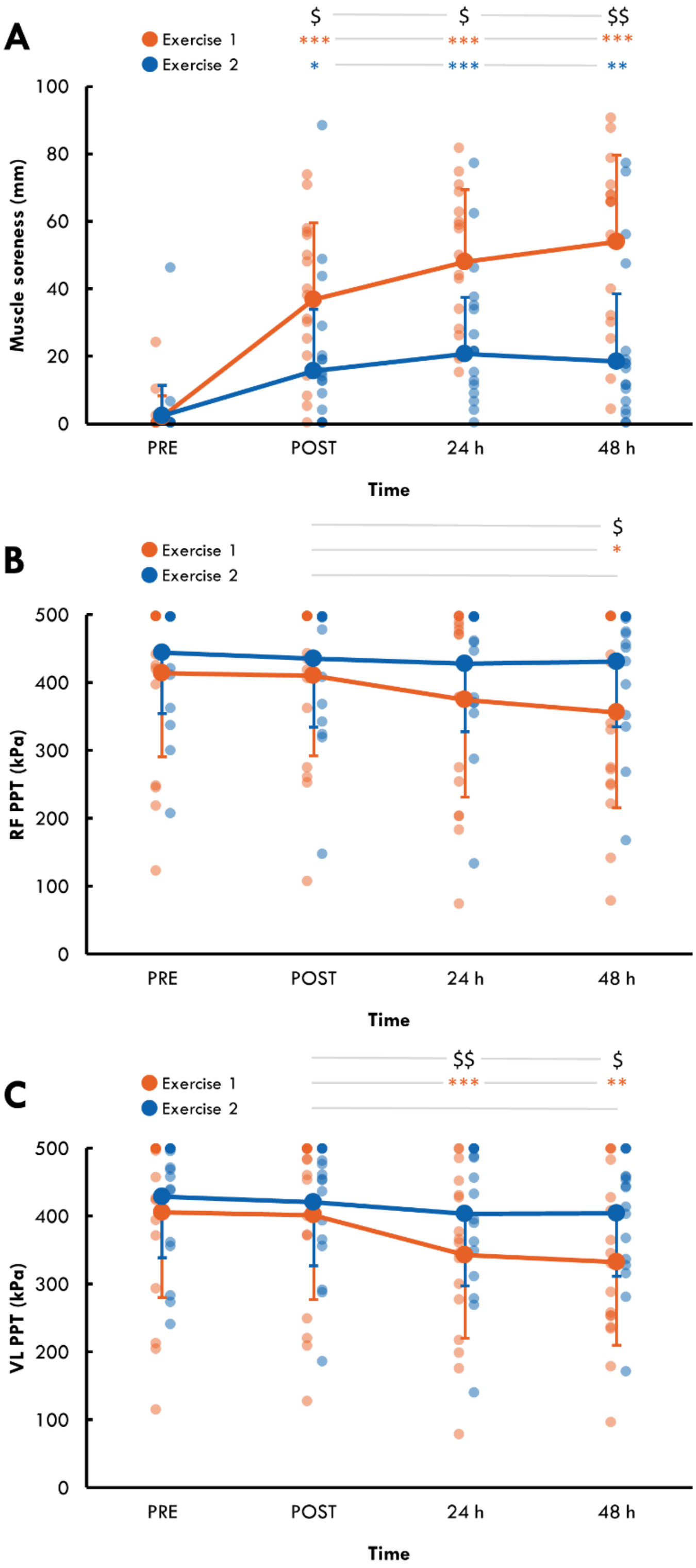

**Figure 5. RF and VL shear elastic modulus (slope) at PRE for exercise 1 and exercise 2.**

$$$: significant difference between exercise 1 and exercise 2 at $p < 0.001$.

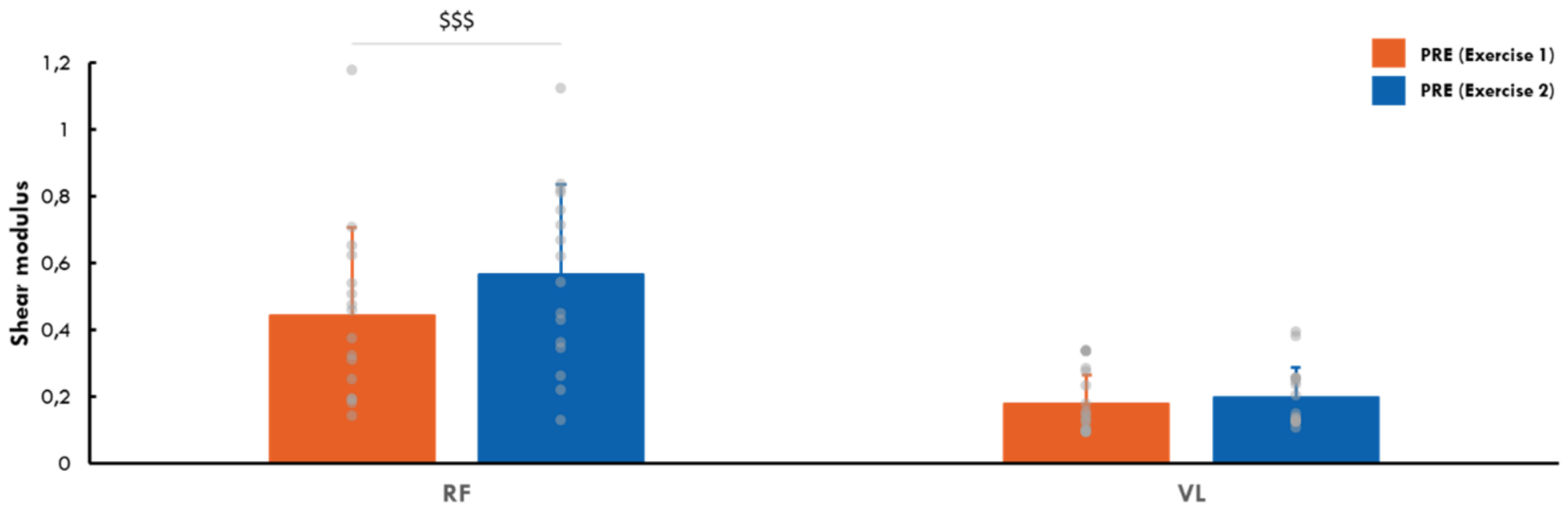

**Figure 6. Time course of (A) *Rectus femoris* (RF) and (B) *Vastus lateralis* (VL) EMG activity of the stance phase (mean of 0-35% of the step duration) and (C) *Biceps femoris* EMG activity of the swing phase (mean of 35-100 % of the stride duration) during exercise 1 (Ex1) and exercise 2 (Ex2).** Significantly different from PRE value: *: $p < 0.05$, **: $p < 0.01$, ***: $p < 0.001$. Significant difference between Ex1 and Ex2: $$$ ($p < 0.001$). Plateau correspond to the first time (% exercise) at which the value is not significantly different from the end-exercise value.

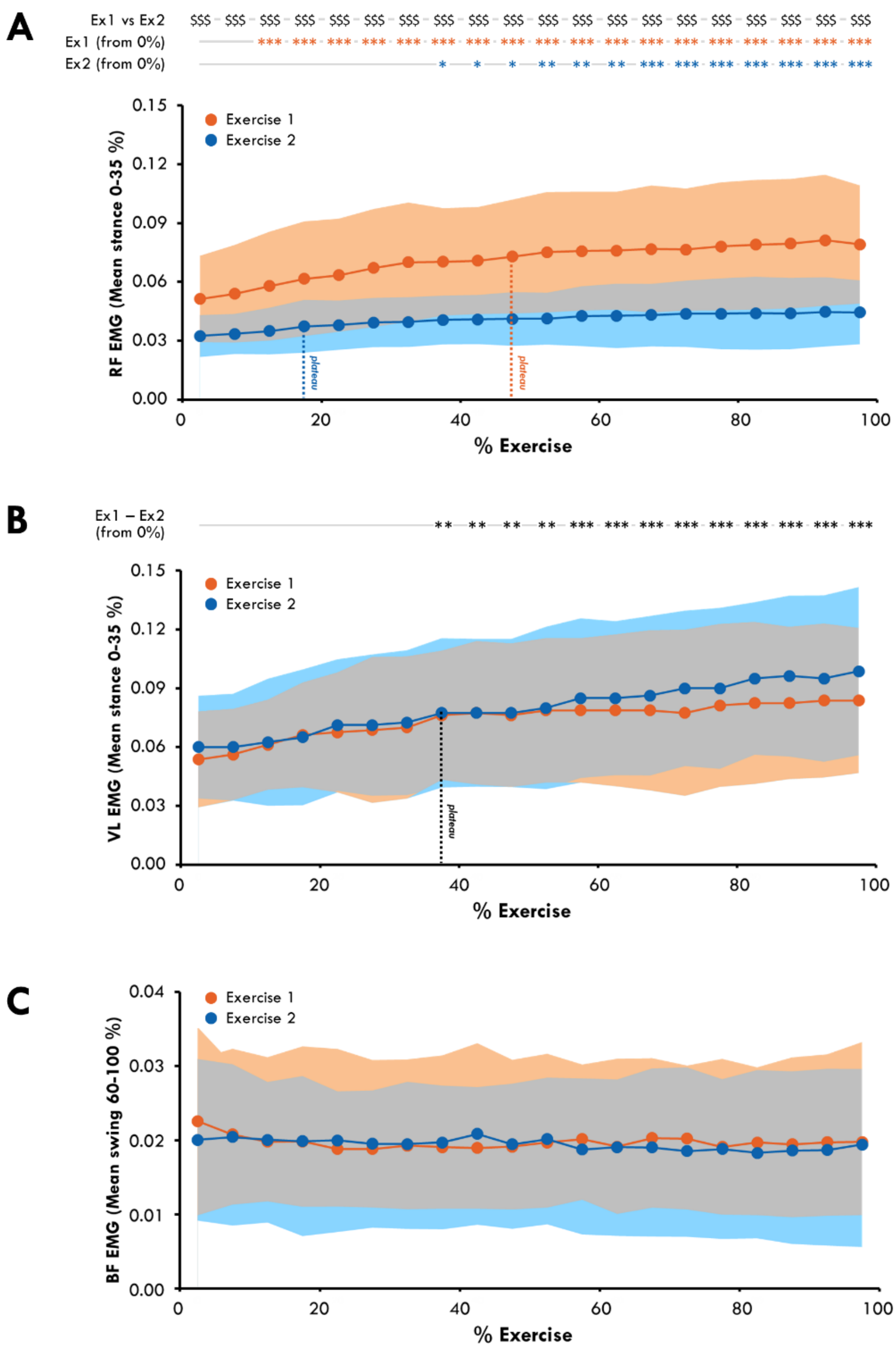

**Figure 7. Time course of (A) cost of walking ($C_w$), (B) heart rate (HR), and (C) rate of perceived exertion (RPE) during exercise 1 (Ex1) and exercise (Ex2).** Significantly different from PRE value: *: $p < 0.05$, **: $p < 0.01$, ***: $p < 0.001$. Significant difference between Ex1 and Ex2: $: $p < 0.05$, $$: $p < 0.01$, $$$: $p < 0.001$. Plateau correspond to the first time (% exercise) at which the value is not significantly different from the end-exercise value.

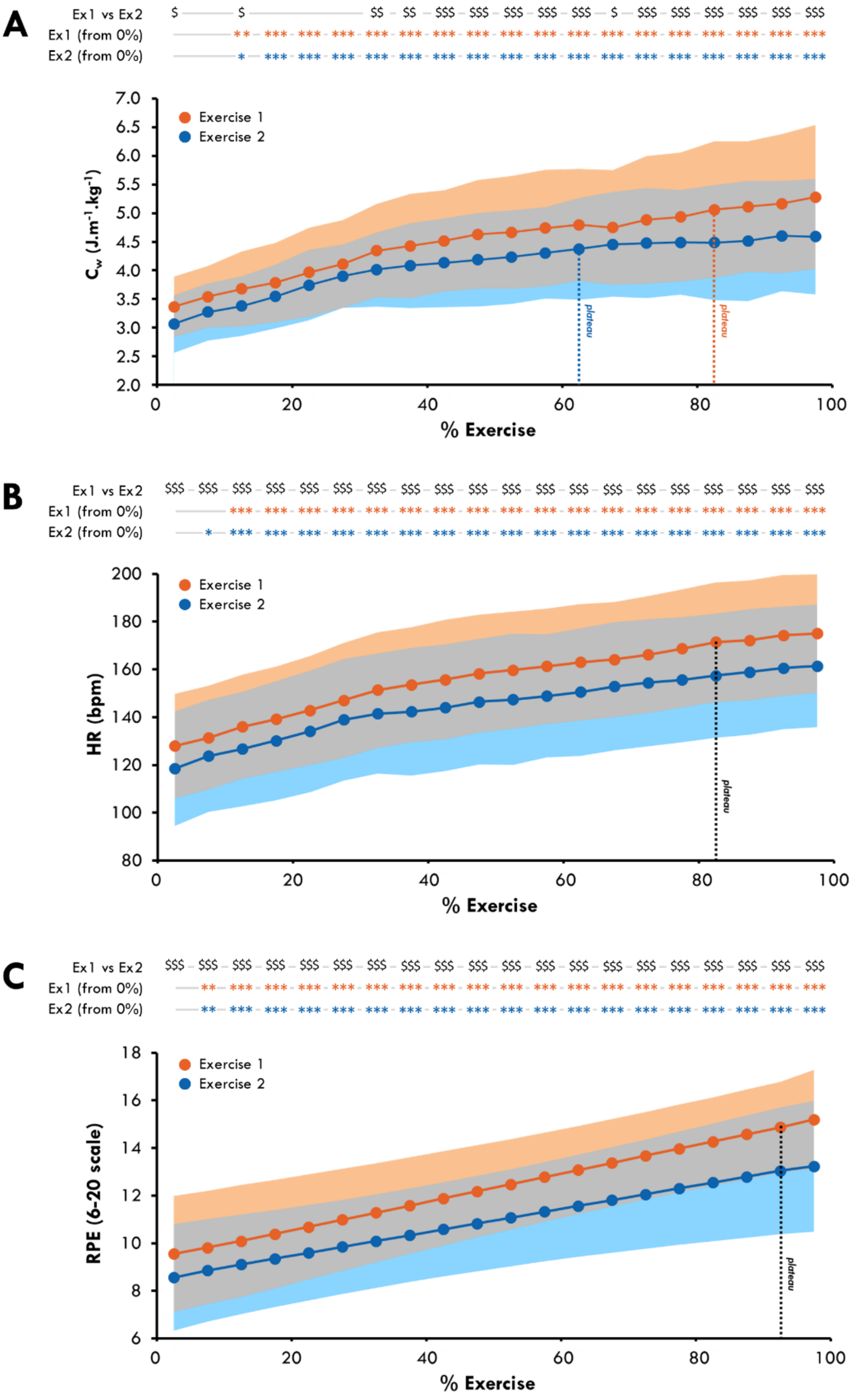